\documentclass[10pt,twocolumn]{article}

\usepackage{graphicx,amssymb,amsmath,amsbsy,bm,amsfonts,epstopdf}

\newtheorem{defn}{Definition}

\newtheorem{prop}{Proposition}
\newtheorem{remark}{Remark}

\def\varp{ {\rm VAR}_p }

\def\eye{{\rm i}}
\def\e{{\rm e}}

\def\bfI{ {\boldsymbol I} }
\def\bfX{ {\boldsymbol X} }
\def\bfS{ {\boldsymbol S} }

\def\bfA{ {\boldsymbol A} }

\def\bfT{ {\boldsymbol T} }

\def\cond{ {\scriptscriptstyle{\bullet}}  }

\def\bfPhi{ {\boldsymbol \Phi} }
\def\bfSigma{ {\boldsymbol \Sigma} }

\def\bfepsilon{ {\boldsymbol \epsilon} }

\def\bfsepsilon{ {\scriptstyle{\bfepsilon}} }

\def\ri{ {\rm i}}
\def\re{ {\rm e} }
\def\eye{ {\rm i}}

\def\dif{ {\rm d} }

\def\cov{\mathop{\rm cov}\nolimits}

\def\tr{ \mbox{tr} }

\def\orthog{ {\perp\!\!\!\!\perp} }

\begin{document}

\markboth{IEEE Transactions on Signal Processing,
[\today]} {Wolstenholme \& Walden: Multiple Hypothesis Testing}

\title{An Efficient Approach to Graphical Modelling of  Time Series}

\author{R.~J.~Wolstenholme  and A.~T.~Walden \thanks{Copyright (c) 2015 IEEE. Personal use of this material is permitted. However, permission to use this material for
any other purposes must be obtained from the IEEE by sending a request to pubs-permissions@ieee.org.
Robert Wolstenholme and Andrew Walden
are both at the Department of Mathematics, Imperial College  London, 180 Queen's Gate,
London SW7 2BZ, UK. 
(e-mail: rjw08@imperial.ac.uk and a.walden@imperial.ac.uk) 
}}

\maketitle

\begin{abstract} 
A method for selecting a graphical model for $p$-vector-valued stationary Gaussian time series was recently proposed by Matsuda and uses the Kullback-Leibler divergence measure to define a test statistic.  This statistic was used in a backward selection procedure, but the algorithm is prohibitively expensive for large $p.$ A high degree of sparsity is not assumed. 
We show that reformulation in terms of a multiple hypothesis test reduces computation time by $O(p^2)$
and simulations support the assertion  that power levels are attained at least as good as those achieved by Matsuda's much slower approach. Moreover, the new scheme is readily parallelizable for even greater speed gains.
\end{abstract} 


\section{Introduction}
There has been much interest in recent years in the construction of graphical models from 
 $p$-vector-valued (or multivariate) stationary time series $\{ {\bfX}_t
\}$ where
${\bfX}_t=[X_{1,t},\ldots,X_{p,t}]^T\in {\mathbb R}^p, \,t\in{\mathbb Z},$
and $^T$ denotes transposition. The purpose of graphical models is to aid visualization of connections between multiple time series: each of the  time series is represented by one vertex
and it is wished to define connections via edges  between the vertices  of the graph. The lack of an edge indicates the lack of a connection between the corresponding series. 

Formally, a graph $G=(V,E)$ consists of vertices $V$ and edges
$E$,  where $E\subset \{(j,k)\in V \times V: j\not=k\}.$ (We are considering simple graphs where there are no loops from a vertex to itself, nor multiple edges between two vertices.)
To represent $\{ {\bfX}_t
\}$ the vertices of the graph correspond to the $p$ individual series $\{ X_{j,t} \}$, so  $V = \{ 1, \ldots, p \}$. 
Edges connect
ordered pairs of distinct vertices. 
Edges $(j,k)\in E$ for which both $(j,k)\in E$ and $(k,j)\in E$ are called
undirected edges. An undirected graph is one with only undirected edges and it only represents interaction between the series. 
An edge $(j,k)$ is called directed if $(j,k)\in E$, with $(k,j)\not\in E.$  A directed graph is one in which all edges are directed and it typically encodes directions of influence or of causation between series. 

In this paper we will consider the modelling only of {\it undirected graphs}. There are $p(p-1)/2$ unordered pairs of vertices for the graph  and  $2^{p(p-1)/2}$ possible distinct graph structures. 
A high degree of sparsity of edges is not assumed. We are interested in both moderate  $p$ and large $p;$ both are practically important and present a challenge to graphical modelling when a high degree of  sparsity is not assumed.

The statistical framework for  graphical modelling of vector-valued time series was begun by  Brillinger \cite{Brillinger96} who considered both directed and undirected graphs.  Two different nonparametric approaches were subsequently developed, by 
Dahlhaus \cite{Dahlhaus00} for undirected graphs, and by Bach and Jordan 
\cite{BachJordan04} for directed graphs.

In the approach of \cite{Dahlhaus00}  the
absence of an edge in the graphical model between series $j$ and $k$ is indicated by the corresponding
partial coherence,  being zero at all frequencies $f.$ The partial coherence is a frequency domain version of the partial correlation coefficient and measures, at a frequency $f,$ the correlation between series $j$ and $k$  when all other series involved are held constant. The partial coherence is
denoted $\gamma^2_{jk\cond\{\setminus jk\}}(f),$ where
$\{\backslash jk\}=\{1 \leq i \leq p : i \neq j,k \},$  and the $_{\cond\{\setminus jk\}}$ terminology indicates that these series are held constant. The assessment of the interaction between series 
$j$ and $k$ thus discounts the indirect effects of the other series. 
Estimated partial coherencies  will include sampling variability and will  never be exactly zero, so that hypothesis testing is required to test edge $(j,k)$ to see if it should be declared to be missing. The problem here is that the partial coherence for edge $(j,k)$ must be zero-tested for every frequency computed:
Dahlhaus \cite{Dahlhaus00}  suggested a test  based simply on  the maximum of the nonparametrically-estimated partial coherence over the frequency range,  but the exact asymptotic null distribution of his test statistic is not known and only  approximations have been used in practice. Nevertheless, this nonparametric approach has seen considerable use 
\cite{FriedDidelez03,Gather_etal02,Timmer_etal00}.

The approach of \cite{BachJordan04} for {\it directed}\/ graphs, while inapplicable here, had as a key component the use of the Kullback-Leibler (KL) divergence  
between stationary processes,  formulated earlier by \cite{Kazakos80}. In this paper we use 
the KL divergence for determining {\it undirected}\/ graphs.

As an alternative to \cite{Dahlhaus00}, it was suggested in 
\cite{Eichler06} and \cite{Songsirietal10} to instead use parametric graphical models, known
as `graphical interaction models'  
which utilise vector autoregressive (VAR) processes to model $\{ \bfX_t \}.$
Here the VAR parameters are constrained by an associated graph; by then ranging
over  {\it all}\/ the $2^{p(p-1)/2}$ possible graphs and (typically low) orders of the VAR model, an information criterion (IC) can be
used to select an appropriate model. 
However such an exhaustive search procedure is only suitable for small $p.$  

As an alternative to such  exhaustive searches, a topology selection scheme which uses
a more efficient approach was given in \cite{SongsiriVandenberghe10}. It uses  penalized maximum likelihood where the penalty term reflects sparsity constraints. For every pair of series, the resulting partial coherence is then subjected to thresholding to determine whether it can be considered to be everywhere null (for determining the missing edges). Having thus determined the missing edges, the graph is determined and constrained parameters can be estimated. By ranging over a small number of possible VAR orders and penalty weights, and computing an IC in each case, the graph giving the minimum value of the IC is selected. Unfortunately, the correct/optimum level to take for the critical thresholding step is unknown in practice.


A fully nonparametric approach to graphical modelling has the advantage of avoiding the possibility of model misspecification that can arise with parametric modelling when addressing real-world data. Indeed,
Matsuda \cite{Matsuda06} proposed the identification of a  graphical model for 
$\{ \bfX_t \}$ based on the use of nonparametrically-estimated Kullback-Leibler (KL) divergence between two graphical models. 
Matsuda's test statistic is simple to compute and its  asymptotic null distribution  is Gaussian.  
It allows to test  whether a particular nested
subgraph is ``correct'' --- in the sense that it contains the true graph --- and thus to determine
if restricting the set of edges poses a real constraint. Matsuda used an iterative
procedure: at each step the null hypothesis that a subgraph with one edge less
is correct is tested. At each such iteration the test therefore has to be carried out as many
times as there are edges remaining in the graph; this is computationally very costly because of the number of test statistics needing to be computed, especially for large $p.$ Moreover, for general non-decomposable graphs the computation of the test
statistic employs  another iterative procedure to satisfy the constraints imposed by the currently
selected graph.

In this paper we introduce a much more efficient approach to identifying the  model --- while still based on Matsuda's test statistic. 
Instead of an iterative procedure, we consider only tests that compare the fully connected or saturated graph (alternative) with
graphs that have exactly one missing edge (null hypothesis). These tests are carried out using the well-known Holm
method for multiple hypothesis testing. The method provides strong familywise error
control which means that the type I error of rejecting any of the tested null hypotheses
falsely does not exceed the specified significance level.
This obviously decreases
the number of tests required as well as the computational burden for evaluating the test
statistics themselves  as iterative fitting algorithms are no longer required.
Indeed, the number of computations for our approach is $O(p^4)$ compared to $O(p^6)$ for Matsuda's implementation. Additionally, in simulations our algorithm achieves power at least as good as that achieved by Matsuda's original and much slower approach.

In Section~\ref{sec:graphmod} we review background ideas in time series graphical modelling (including the concept of a {\it correct}\/ graph).  Section~\ref{sec.gmod} summarizes the construction of Matsuda's test statistic and gives a worked example showing how it is used in his backward stepwise selection procedure. In Section~\ref{sec:newalg} we describe our much more computationally efficient multiple hypothesis test (MHT) employing Matsuda's test statistic. The computational efficiencies of the two approaches are contrasted in Section~\ref{sec:efficiency}, justifying the $O(p^2)$ improvement for the MHT algorithm, empirically illustrated in   
Section~\ref{subsec:timingsone}. Statistical powers are compared for the two algorithms in Section~\ref{sec:power}, and the MHT algorithm is seen to do at least as well as Matsuda's algorithm. That the MHT algorithm performs well for higher-dimensional models (large $p$),  and  is readily parallelizable for even greater speed gains, is shown in Section~\ref{sec:highdims}. The methodology is satisfactorily applied to $p=10$ EEG data in Section~\ref{sec:EEG}. Concluding comments are provided in Section~\ref{sec:summary}.

\section{Graphs and VAR Models}\label{sec:graphmod}
Throughout the paper, for a matrix ${\bfA}$, $A_{jk}$ refers to the $(j,k)$th element of ${\bfA}$ and $A^{jk}$ refers to the $(j,k)$th element of ${\bfA}^{-1}$, unless otherwise stated. Without loss of generality $\{ {\bfX}_t
\}$ is taken to have a mean of zero.

\subsection{Time Series Graphical Models}

The edges between the vertices represent partial correlation between two series, i.e., there is no connection between nodes $j$ \& $k$ if and only if $X_j$ and $X_k$ are partially uncorrelated given $X_{\{\backslash jk\}}.$ 
To be precise, we remove the linear effects of $X_{\{\backslash jk\}}$ from $X_j$ to obtain the $j$th residual series defined as
$
\nu_{j,t}=  X_{j,t}- \sum_{v \in \{\setminus jk\}} \sum_{u} a_{jv,u} X_{v,t-u}, 
$
where the $p-2$ filters $\{ a_{jv,u}, u \in {\Bbb Z}\}$ give the minimum mean square
prediction error. The $k$th residual series is defined likewise.
The sequence $s_{\nu_j\nu_k,\tau}=\cov\{\nu_{j,t+\tau},\nu_{k,t}\}, \tau \in {\Bbb Z},$  is called the partial
cross-covariance sequence and the 
two residual series are uncorrelated if it is everywhere zero. 
If $X_j$ and $X_k$ are partially uncorrelated  we write $X_j \orthog X_k | X_{\{\backslash jk\}}$.
Let
$
(j,k) \not\in E \Longleftrightarrow X_{j} \orthog X_{k} | X_{\{\setminus jk\}}.
$ Then
$G$ is called a partial correlation graph. 
For Gaussian time series a null partial correlation equates to independence between the $j$th and $k$th conditioned series, and in this case we have a conditional independence graph.  

The Fourier transform of the partial
cross-covariance sequence is the partial cross-spectral density function, denoted $S_{jk\cond(\setminus jk)}(f).$ 
The partial coherence, for $-1/2 \leq f < 1/2$, is defined as 
$$
\gamma^2_{jk\cond(\setminus jk)}(f) = {|S_{jk\cond(\setminus jk)}(f)|^2 \over {S_{jj\cond(\setminus jk)}(f) S_{kk\cond(\setminus jk)}(f)} }.
$$
Since
$S_{jk\cond(\setminus jk)}(f)\equiv 0 \,\,\hbox{for
all}\,\,-1/2 \leq f < 1/2\Longleftrightarrow s_{\nu_j \nu_k,\tau}=0 \,\,\hbox{for
all}\,\,\tau\in\Bbb{Z}$ 
we see that 
\begin{equation*}
(j,k) \not\in E \Longleftrightarrow 
S_{jk\cond(\setminus jk)}(\cdot) \equiv 0 \Longleftrightarrow \gamma^2_{jk\cond(\setminus
jk)}(\cdot)\equiv 0.
\end{equation*}

Let $\bfS(f)$ denote the spectral matrix of $\{ \bfX_t\}$ at frequency $f,$
assumed to exist and be of full rank. 
Denoting the $(j,k)$th element of $\bfS^{-1}$ by $S^{jk}(f),$
the partial coherence can be expressed as,
(e.g., \cite{Dahlhaus00}),
$
\gamma^2_{jk\cond\{\setminus jk\}}(f) 
= {|S^{jk}(f)|^2 / [S^{jj}(f) S^{kk}(f)] },
$
and therefore
\begin{equation*} 
(j,k) \notin E  \iff S^{jk}(f) = 0, \qquad  -1/2 \leq f < 1/2.
\end{equation*}
i.e., if $X_j$ and $X_k$ are partially uncorrelated then there is a zero in the corresponding entry of the inverse spectral matrix  \cite{Dahlhaus00}.
(Partial correlation graphical models for time series are  undirected as $(j,k) \notin E \iff (k,j) \notin E$. )

\subsection{Correct Graphs}
An important concept in what follows is that of a {\sl correct graph}.
Such graphs can be used  to identify the underlying graphical model for multivariate time series. The following definition is a slightly clarified version of that in  \cite{Matsuda06}.

\begin{defn}\label{def:one}
If $(V,E)$ is the true graphical model for $\{ {\bfX}_t \}$, then $(V,E')$ is correct for $(V,E)$, if 
\begin{equation}\label{eq:correct}
S^{jk}(f) = 0, \quad (j,k) \notin E'\quad\mbox{and} -1/2 \leq f < 1/2.
\end{equation}
\end{defn}
\smallskip
Note that by this definition, if an edge is missing in $E'$ it must also be absent in $E$ for $(V,E')$ to be correct.  A correct graph $(V,E^{'})$, when imposed on top of $(V,E)$, will completely cover all its edges as $E \subseteq E^{'}$. Also,
the fully saturated graph --- containing all edges between vertices --- is correct for any graphical model.

\begin{figure}[t!]
\begin{center}
\includegraphics[scale=1]{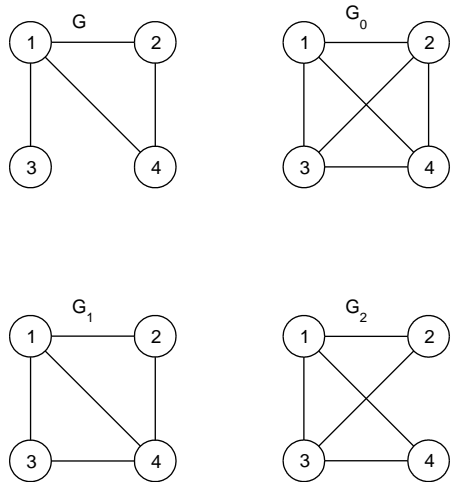}
\end{center}
\caption{Illustration of the concept of a correct graph. $G$ is the true graph. As explained in the text, $G_0$ and $G_1$ are correct for $G$ while $G_2$ is not.}
\label{fig:correct_graph}
\end{figure}

By way of an example, let 
$G=(V,E)$ in Fig.~\ref{fig:correct_graph} be the true graphical model.
Then the fully saturated graph
$G_0=(V,E_0)$ completely covers $G$ and is correct for $G.$
Likewise, $G_1=(V,E^{'})$ completely covers $G$ and is correct for $G.$
However, when $G_2=(V,E^{''})$ 
is imposed over $G$ the  edge between $\{X_{2,t}\}$ and $\{X_{4,t}\}$ in $G$ is not covered. So $(2,4) \notin E^{''}$ but $(2,4) \in E$. Therefore $E \not \subseteq E^{''}$ and $G_2$ is not a correct graph for $G.$

It should be emphasized that we use the phrasing
``$(V,E)$ is the {\it true} graphical model for $\{ {\bfX}_t \}$'' and reserve the use of the word {\it correct} for 
the special context of Definition~\ref{def:one}.

\subsection{VAR Models}\label{subsec:Vars}

Here we give a very brief summary of some relevant results on VAR processes, useful for understanding ideas in our simulation examples such as ``jointly influencing.'' We stress however that the methodology discussed in the paper is  more widely applicable.

$\{ {\bfX}_t \}$ is a real-valued zero mean $p$-vector-valued
autoregressive process of order $\ell$, or $\varp(\ell),$ 
if it is of the form
$
{\bfX}_t= \sum_{u=1}^{\ell} \bfPhi_u{\bfX}_{t-u}+ \bfepsilon_t,
$
where the $\{\bfPhi_u\}$ are $p\times p$ coefficient matrices, and
$\bfepsilon_t=[\epsilon_{1,t},\ldots,\epsilon_{p,t}]^T$ is a $p$-vector-valued white noise
process with a mean vector of zero and covariance matrix $\bfSigma_{\bfsepsilon}.$ 
If 
$
\det\{ {\bfI}_p-\sum_{u=1}^{\ell} \bfPhi_u z^u \}\not=0\,\,\mbox{for all}\,\, |z|\leq 1,
$
where $ {\bfI}_p$ is a $p\times p$ identity matrix, then the process is stationary 
\cite[p.~25]{Lutkepohl06}.
We define 
$
\bfPhi(f)\equiv  - \sum_{u=0}^{\ell} \bfPhi_u \e^{-\eye 2\pi fu }
$
and $\bfPhi_0 \equiv -\bfI_p.$

Let $\Phi_{ij,u}$ be the $(i,j)$th element of $\bfPhi_u$ where we are interested in the case $i\not =j.$ Then $\Phi_{ij,u}$ is said to be the {\sl influence} from $X_{j,t-u}$ on $X_{i,t}$ \cite{Dahlhaus00}. There is no  influence from component $j$ on $i$ if $\Phi_{ij,u}=0, u=1,\ldots,\ell,$
so that $\Phi_{ij}(\cdot)=0.$

$
{\bfS}(f)=  {\bfPhi}^{-1}(f)  \bfSigma_{\bfsepsilon} [{\bfPhi}^{-1}(f)]^H,
 -1/2 \leq f < 1/2,
$
is the spectral matrix for $\{ {\bfX}_t \}$ 
where $^H$ denotes conjugate transpose.
Then
$
{\bfS}^{-1}(f)=  {\bfPhi}^{H}(f) \,\bfSigma_{\bfsepsilon}^{-1} \, {\bfPhi}(f),
\, -1/2 \leq f < 1/2.
$
If $\bfSigma_{\bfsepsilon}=\sigma_\bfepsilon^2 \bfI_p$
it follows \cite{Dahlhaus00} that if the $j$th and $k$th series do not jointly influence
another series $i\not=j,k$ (i.e., $\Phi_{ij}(\cdot)=0$ and/or $\Phi_{ik}(\cdot)=0$), then the $j$th and $k$th series will be partially uncorrelated if and only if $\Phi_{jk}(\cdot)=0$ and $\Phi_{kj}(\cdot)=0.$

Later we will make use of the VAR$_5(1)$ model
\begin{equation} \label{varModel}
{\bfX_{t}} = {\bf\Phi}_{1} {\bfX_{t-1}} + {\bfepsilon_{t}}
\end{equation}
where $\bfepsilon_t  \sim {\cal N}_5({\bf 0},\bfSigma_{\bfsepsilon}),$
 the $5$-dimensional Gaussian distribution with mean ${\bf 0}$ and covariance matrix $\bfSigma_{\bfsepsilon}.$ 

For testing and illustration purposes we will make use of several models,
named as follows:
\begin{itemize}
\item[]{Model A:}
Here $\bfSigma_{\bfsepsilon}=\bfI_5$
and
\begin{equation}\label{eq:modelone}
 \boldsymbol{\Phi}_1 = \left [ \begin{array}{ccccc}
0.2 & 0 & -0.1 & 0 & -0.5 \\
0.4 & -0.2 & 0 & 0.2 & 0 \\
-0.2 & 0 & 0.3 & 0 & 0.1 \\
0.3 & 0.1 & 0 & 0.3 & 0 \\
0 & 0 & 0 & 0.5 & 0.2 \end{array} \right ].
\end{equation}
By inspection of $\boldsymbol{\Phi}_1$, 
 we see that the set of missing edges is  $ \{ (2,3), (2,5), (3,4)\}.$ 
\item[]{Model B:}
(Matsuda \cite{Matsuda06}). Here $\bfSigma_{\bfsepsilon}=\bfI_5$ and 
\begin{equation} \bfPhi_1 = \left[ \begin{matrix}
0.2 & 0 & 0.3 & 0 & 0.3 \\
0.3 & -0.2 & x & 0 & 0 \\
0.2 & x & 0.3 & 0 & 0 \\
0.2 & 0.3 & 0 & 0.3 & 0 \\
0.2 & 0 & 0.2 & 0.2 & 0.2 \end{matrix} \right ],\label{eq:MatPhi}
\end{equation}
and we consider the cases $x=0$ and $ 0.1,$ as used in \cite{Matsuda06}.
For $x=0$ the set of missing edges in our model is $\{ (2,3), (2,5) \},$ 
where we note that although entries $(3,4)$ and $(4,3)$ are both zero, neither entry $(5,3)$ nor
$(5,4)$ are zero so that series 3 and 4 jointly influence the 5th,  and therefore  edge (3,4) is not missing.
When $x = 0.1$, the set of missing edges is simply $\{(2,5) \}$. 
\item[]{Model C:} This consists of $\bfPhi_1$ of the form (\ref{eq:MatPhi}) with 
$x=0$ but now with $\bfSigma^{-1}_{\bfsepsilon}=\bfI_5$ except that entries 
$(1,2)$ and $(2,1)$ of $\bfSigma^{-1}$ are equal to 0.5. As a result of these two off-diagonal entries being non-zero,
instead of missing edges  $\{ (2,3), (2,5) \}$ only $(2,3)$ is missing.


\end{itemize}

\section{Test Statistic} \label{sec.gmod}

\subsection{Test for Missing Edges}

Given $\{\bfX_{t}\}$ with graph $(V,E)$ and spectral matrix ${\bfS}(f)$, consider graph $(V,E')$ and matrix $\bfT(f)$ satisfying
\begin{equation}\label{eqn.spectralestimate}
T_{jk}(f)=S_{jk}(f), \,(j,k) \in E'; 
\,\,
T^{jk}(f)=0, \,  (j,k) \notin E'.
\end{equation}
Unique existence of $\bfT(f)$ is shown in \cite[Lemma~7]{Matsudaetal06}. 
\begin{prop}\label{prop.correct}
\cite[p.~401]{Matsuda06} Given graph $(V,E')$, if $\bfT(f)$ satisfies the constraints in (\ref{eqn.spectralestimate}) then
$(V,E')$ is correct for $(V, E)$ if and only if $\bfS(f) = \bfT(f).$
\end{prop}
%

The result in Proposition~\ref{prop.correct} can be used to determine whether graph $(V,E_{2})$ is correct, given $(V,E_{1})$ is correct, where $E_{2} \subseteq E_{1}$. With $(V,E_{1})$ assumed correct we have $\boldsymbol{T}_{1}(f) = \boldsymbol{S}(f)$. If we calculate estimators $\hat{\boldsymbol{T}}_{1}(f)$, $\hat{\boldsymbol{T}}_{2}(f)$ using observed data, then intuitively a large difference between them suggests $\hat{\boldsymbol{T}}_{2}(f) \neq \hat{\boldsymbol{T}}_{1}(f) \approx\boldsymbol{S}(f)$ and by 
Proposition~\ref{prop.correct}, $(V,E_{2})$ would be deemed incorrect.

Assuming $(V,E_{1})$ is correct,  a test can be constructed
between a null ($H_0$) and alternative ($H_A$) hypothesis:
$$
H_{0}: (V,E_{2}) \,\, \mbox{is correct} \,\, \mbox{vs}\,\, H_{A}: (V,E_{2}) \,\,\mbox{is incorrect}$$
where a measure of divergence between $\hat{\bfT}_{1}(f)$ and $\hat{\bfT}_{2}(f)$ is used to build the test statistic.

For example, suppose
we want to determine whether two series  are partially uncorrelated, or in fact simply uncorrelated in this case. Define
\begin{equation} \hat{\bfT_{1}}(f) = \left [ \begin{array}{cc}
\hat{S}_{11}(f) & \hat{S}_{12}(f)\\
\hat{S}_{12}^*(f) & \hat{S}_{22}(f) \end{array} \right ],\label{eq:formT1}
\end{equation}
the estimated spectral matrix.
With $(V,E_{1})$ being the fully saturated model, with the two vertices connected, we can test  against $(V,E_{2})$, the model where the vertices aren't connected. The matrix satisfying  (\ref{eqn.spectralestimate}) for $ (V,E_{2})$ is then
\begin{equation} \hat{\boldsymbol{T}}_{2}(f) =
\left [ \begin{array}{cc}
\hat{S}_{11}(f) & 0\\
0 & \hat{S}_{22}(f) \end{array} \right ].\label{eq:formT2}
\end{equation}
%

\subsection{Spectral Estimator}
 Given vector observations ${\bfX_{0}}, \ldots, {\bfX_{N-1}}$, the matrix periodogram estimator $\hat{{\bfS}}^{(P)}(f)$ of $\bfS(f)$ takes the form
$\hat{{\bfS}}^{(P)}(f) = \boldsymbol{W}(f)\boldsymbol{W}^{H}(f),
$
where
$
\boldsymbol{W}(f) =  \sum_{t=0}^{N-1} {\bfX_{t}} \re^{-\ri 2 \pi ft}/\surd N.
$
$\hat{{\bfS}}^{(P)}(f)$ has unit periodicity.
Let $f_{j} = j/N$, the $j$th Fourier frequency, then given a symmetric positive weight sequence $\{w_{k}\}$ for $k=-M, \ldots, M$, 
with $\sum w_{k}=1,$ the frequency-averaged periodogram is
\begin{equation} \label{eqn.wtdpdgm}
\hat{{\bfS}}(f_{j}) = 
\sum_{k=-M}^{M} w_{k} \hat{{\bfS}}^{(P)}(f_{j-k}).
\end{equation}
This estimator was used by Matsuda \cite{Matsuda06} in the derivation of his test statistic. It is necessary and sufficient
for $\hat{\bfS}(f)$ to be non-singular that $2M+1\geq p,$ i.e., we have $p$ or more non-zero values in our weight sequence, e.g., \cite[p.~3007]{Fiecasetal10}. For consistency of the spectral estimator we require $M, N \rightarrow \infty$ such that $M/N \rightarrow 0;$ for the finite sample sizes used in  practice we would expect $M>>p.$ $M$ can be chosen using, for example, the method of `window closing' \cite{Priestley81} 
or by cross-validation \cite{Matsuda06}.

\subsection{Construction of Test Statistic}

 Estimators $\hat{\bfT}_{1}(f)$ and $\hat{\bfT}_{2}(f)$ can be found  by applying the constraints in (\ref{eqn.spectralestimate}) to $\hat{\boldsymbol{S}}(f_{j})$ in (\ref{eqn.wtdpdgm}); the recursion of \cite{WermuthScheidt77}
is used for this purpose 
along with a result from \cite{SpeedKiiveri86} which justifies convergence --- see \cite[p.~403]{Matsuda06}.

To measure the difference between $\hat{\bfT}_{1}(f)$ and $\hat{\bfT}_{2}(f)$ Matsuda \cite{Matsuda06} used the  estimated Kullback-Leibler divergence, $eKL(\bfT_{1},\bfT_{2}).$
With $N$ assumed even this is  
\begin{eqnarray*}
&&\frac{1}{N} \sum_{j=1}^{N/2} \!\left [ \tr\{ \hat{\bfT}_{1}(f_{j})\hat{\bfT}_{2}^{-1}(f_{j}) \}  \right.\\
&&\left.- \log\det\{\hat{\bfT}_{1}(f_{j})\hat{\bfT}_{2}^{-1}(f_{j})\}-p \right ].
\end{eqnarray*}
Under the following assumptions,  Matsuda derived a statistic based on $eKL(\bfT_{1},\bfT_{2})$ which has 
an asymptotically standard normal, ${\cal N}(0,1),$ statistic:
\begin{enumerate}
\item $\{ \bfX_t \}$ is a $p$-vector-valued Gaussian stationary process.
  \item $\bfS(f)$ is positive definite for $|f| \leq 1/2$.
  \item $S_{jk}(f)$ is twice continuously differentiable for $j,k = 1, \ldots, p$ and $-1/2 \leq f < 1/2$.
  \item $M$ = $O(N^{\beta})$ ($M$ is at most of order $N^\beta$) for $1/2 < \beta < 3/4$ and the weight sequence $\{w_{k}\}$ 
is of the form
$ 
w_{k} = u\left(\frac{k}{2M}\right), \,\, k = -M,\ldots, M,
$
where $u(\cdot)$ is a continuous even function on $[-1/2,1/2].$
\end{enumerate}

Matsuda \cite{Matsuda06} defined the test statistic $Z_{N}(\bfT_{1},\bfT_{2})$ as
\begin{equation} \label{eqn.testStat}
\left [ \frac{2MN}{D_{u}(m_{2} - m_{1})} \right ]^{1/2} \left [ eKL(\bfT_{1}, \bfT_{2}) - \frac{C_{u}(m_{2} - m_{1})}{2M} \right ]
\end{equation}
where $m_{i} = \# \{ (j,k) : (j,k) \notin E_{i},\, j<k \},$ (the number of missing edges in the model), and $C_u, D_u$ are constants with values determined by $u(\cdot),$ see \cite{Matsuda06}.
Given assumptions 1-4 it follows that \cite{Matsuda06}
\begin{itemize}
\item Under $H_{0},$
\begin{equation}\label{eq:sn}
Z_{N}(\bfT_{1},\bfT_{2}) \rightarrow {\cal N}(0,1) \,\, \mbox{as}\,\, N \rightarrow \infty
\end{equation}
\item Under $H_{A}, Z_{N}(\bfT_{1},\bfT_{2})$ takes the form
\begin{equation}\label{eq:underH1}
\left [ \frac{2MN}{D_{u}(m_{2} - m_{1})} \right ]^{1/2} KL(\bfS,\bfT_{2})  + o_{p}([MN]^{1/2})
\end{equation}
where $\bfS$ is the true spectral matrix, $KL(\cdot,\cdot)$ denotes the true
Kullback-Leibler divergence, and $o_{p}([MN]^{1/2})$ denotes a term of smaller order in probability than $[MN]^{1/2}.$
\end{itemize}

Under $H_{A}$, the dominant term of the test statistic, the divergence, is positive and it therefore has a one-sided critical region. So for values of the statistic greater than a critical level, $H_{0}$ is rejected in favour of $H_{A}$.
Also from (\ref{eq:underH1}) the statistic diverges to infinity at rate $[MN]^{1/2}$ under $H_{A},$ so that the test can be more powerful than other standard tests which diverge at the rate $N^{1/2}$ \cite{Matsuda06}.

\begin{remark}
We draw attention to the fact that Matsuda's statistical results 
 assume  that the processes involved are Gaussian. He considered 
\cite[p.~407]{Matsuda06} that this might not be a necessity, but presently this is an open question. Bach and Jordan 
\cite{BachJordan04} also assumed Gaussianity in their study for directed graphs.

\end{remark}

\subsection{Matsuda's Algortihm}
Matsuda \cite{Matsuda06} used the test statistic (\ref{eqn.testStat}) and the recursion in \cite{WermuthScheidt77} in a backward stepwise selection algorithm to identify the best graphical model for $ \{ \boldsymbol{X}_{t} \}$. 
Start by setting $(V,E_{0})$ equal to the fully saturated graph with no missing edges and choose significance level $\alpha$. Set $k=0$ and begin: \begin{enumerate}
\item Let $(V,E_{k+1}^{1}), (V,E_{k+1}^{2}), \ldots, (V,E_{k+1}^{L_k})$ be the $L_k$ distinct graphs with one more missing edge than $(V,E_{k})$. Calculate the test statistics
$$Z_{N}^{i}=Z_{N}(\bfT_{k},\bfT_{k+1}^{i}), \quad i = 1, \dots, L_k,
$$
with $\bfT_{k+1}^{i}$  the statistic corresponding to  model $(V,E_{k+1}^{i}).$
\item With $\Phi(\cdot)$ denoting the standard Gaussian distribution function, find $C_k(\alpha)$ satisfying
\begin{equation}\label{eq:alphaMat}
C_k(\alpha) = \Phi^{-1}((1-\alpha)^{1/L_k})
\end{equation}
and if for all $i$, $Z_{N}^{i} > C_k(\alpha)$, then stop the procedure and select $(V,E_{k})$ as the graphical model for $ \{ \boldsymbol{X}_{t} \}$. Otherwise, set $(V,E_{k+1})=(V,E_{k+1}^{j})$ where $Z_{N}^{j}$ is the smallest statistic calculated.
\item Set $k=k+1$ and loop back to step 1.
\end{enumerate}

Under the assumption that all $Z_{N}^{i}$ are standard Gaussian --- which they will be asymptotically if $(V,E_{k+1}^{i})$ is a correct graph --- the result
\begin{eqnarray*}
P\left\{\bigcap_{i=1}^{L_k}(Z_{N}^{i} \leq C_k(\alpha))\right\} &\geq& \prod_{i=1}^{L_k} P\{Z_{n}^{i} \leq C_k(\alpha)\}\\
 &=& (1 - \alpha)
\end{eqnarray*}
means that under the hypothesis that {\it all}\/ $(V,E_{k+1}^{i})$ are correct, the type I error rate is asymptotically less than $\alpha$ and the critical region is conservative \cite[p.~404]{Matsuda06}.
\begin{remark}
%
Perhaps a more intuitive definition  for the type I error rate, which we use later,  would be the probability of not removing an edge when $(V,E_{k+1}^{i})$ is correct, i.e., it should have been removed. This is because we know the distribution of $Z_{N}^{i}$ when $(V,E_{k+1}^{i}$) is correct, so this error rate can be calculated.  The error rate used in the stepwise selection is only relevant in terms of the tests carried out at each step. It is unclear how it is related to the overall properties of the procedure \cite[p.~158]{Edwards00}.
\end{remark}

\subsection{Worked Example} \label{sec.workex}

The weight function chosen is $w_{k} = \cos(\pi k/2M),\,k=-M, \ldots, M$ with $M=64.$ Numerical evaluation of  $C_u$ and $D_u$ when $u(x)=\cos(\pi x)$ gives $C_{u}=0.617$ and $D_{u}=0.446$. 
We consider Model~A of Section~\ref{subsec:Vars}
with
missing edges   $ \{ (2,3), (2,5), (3,4)\}.$ 
With $N=1024$ for simulations of the VAR process, we ran Matsuda's algorithm with significance level $\alpha = 0.05$.

Let $(V,E_{0})$ be the completely saturated graph.  The test statistics 
$Z_N(T_k,T_{k+1}^i)$ for the potential models and the critical levels 
$C_k(0.05)$ at which they are tested are given in Table~\ref{tab:MatsudaEx}.
The steps are interpreted as follows:
\begin{table}[t!]
\begin{center}
\begin{tabular}{|c|c|c|c|c|}
\hline 
Edge & $k=0$ & $k=1$ & $k=2$ & $k=3$
\tabularnewline
\hline 
\hline 
(1,2) & 53.71 & 54.03 &  57.02 & 67.63\\
(1,3) & 12.72 & 14.62 & 17.55 & 17.54\\
(1,4) & 22.25 & 24.14 & 24.14 & 23.71\\
(1,5) & 67.92 & 68.96 & 70.12 & 79.62\\
(2,3) & 0.54 & {\bf 0.20} & --- & --- \\
(2,4) & 18.16 & 17.82 & 17.82 & 22.41\\
(2,5) & 1.89 & 1.90 & {\bf 1.94} & --- \\
(3,4) & {\bf 0.21} & --- & --- & --- \\
(3,5) & 5.86 & 5.29 & 5.50 & 5.49\\
(4,5) & 73.17 & 72.60 & 72.60 & 77.23
\tabularnewline
\hline
\hline
$C_k(0.05)$ & 2.53 & 2.49 & 2.44 & 2.39\\
\hline
\end{tabular}
\end{center}
\caption{Test statistics $Z_N(T_k,T_{k+1}^i)$ and critical levels $C_k(0.05) $ for Matsuda's algorithm}
\label{tab:MatsudaEx}
\end{table}
\begin{description}
\item{$k=0:$}
Not all test statistics are above the critical level, so the process does not stop; $(V,E_{1})$ is set to the graph with the edge $\{(3,4)\}$ missing as this had the lowest corresponding test statistic.
\item{$k=1:$} Likewise
$(V,E_{2})$ is set to the graph with the edges $\{(2,3),(3,4)\}$ missing as $(2,3)$ had the lowest corresponding test statistic.
\item{$k=2:$} Likewise
$(V,E_{3})$ is set to the graph with the edges $\{(2,3),(2,5),(3,4)\}$ missing as $(2,5)$ had the lowest corresponding test statistic.
\item{$k=3:$}
At this step all the statistics are above $C_3(0.05);$ we stop the process here and take $(V,E_{3})$ as the estimated graph. 
\end{description}

This procedure gave the true final graph for the model.

\section{An Efficient Testing Procedure}\label{sec:newalg}
\subsection{Multiple Hypothesis Testing}
We now introduce a new and much more efficient approach for identifying the true graphical model for $\{\boldsymbol{X}_{t}\}$. While still based on the test statistic defined in (\ref{eqn.testStat}), our method doesn't update at each iteration. Essentially, we carry out Matsuda's method only for $k=0$, taking $(V,E_{0})$ as the fully saturated graph. 
If the value of the statistic $Z_{N}(\bfT_{0},\bfT_{1}^{i})$ 
corresponding to graph $(V,E_{1}^{i})$ is
below an appropriate critical level, it is deemed a correct graph and the missing edge $i$ should also be missing in the estimated graphical model. We construct our estimated model by removing insignificant edges via a MHT.

Our null hypotheses are of the form $H_i: (V, E_1^i)$  is correct. The alternative hypothesis in each case is the fully connected or saturated graph.
Each test is thus concerned with whether an edge exists between two vertices specified by the value of $i$.

\begin{prop}
If the graph $(V,E_{1}^{i})$ is correct for edges corresponding to $i=i_{1}, \dots ,i_{s}$ and incorrect for all others, then the graphical model $(V,E)$ for $\{\boldsymbol{X}_{t}\}$ is the graph with only edges $\{i_{1}, \dots ,i_{s}\}$ missing.
\end{prop}

{\it Proof:\/}
If graph $(V,E_{1}^{i})$ is correct and $i$ corresponds to the edge $(j,k)$, then by definition $S^{jk}(f) = 0$ for $-1/2 \leq f < 1/2$ where $\bfS(f)$ is the spectral matrix of the true graphical model. This means that edge $(j,k)$ must also be missing in $(V,E)$ and this is the case for all $i=i_{1}, \ldots, i_{s}$. 
Conversely, if $(V,E_{1}^{i})$ is incorrect,  ${S}^{jk}(f) \neq 0$ and $(j,k)$ must necessarily be in $(V,E)$, hence the result.
\hfill$\square$
\medskip

We can list the $L=p(p-1)/2$ hypotheses in an obvious way:
\begin{eqnarray*}
H_{1}:&&  (V, E_1^1)\,\, \mbox{is correct;}\,\,  (1,2) \not\in E \\
&&\qquad\vdots   \\
H_{p-1}:&&  (V, E_1^{p-1})\,\, \mbox{is correct;}\,\,  (1,p) \not\in E\\
H_{p}: && (V, E_1^p)\,\, \mbox{is correct;}\,\,  (2,3) \not\in E \\
&&\qquad\vdots   \\
H_{L}:&& (V, E_1^L)\,\, \mbox{is correct;}\,\,  (p-1,p) \not\in E.
\end{eqnarray*}
Multiple hypothesis testing may be addressed via the maximin stepdown procedure 
\cite[Sec.~9.2]{LehmannRomano05}.
With $Z_{N}^{i}\equiv Z_{N}(\bfT_{0},\bfT_{1}^{i})$ for $i = 1, \ldots, L$ and  {\it ordered} test statistics
$Z_{N}^{(1)} \leq \cdots \leq Z_{N}^{(L)}$ the corresponding hypotheses $H_{(1)}, \ldots, H_{(L)}$ can be tested using the maximin stepdown procedure:
\begin{itemize}
\item {\em Step 1}: if $Z_N^{(L)}<C_L$, accept $H_{1},\ldots,H_{L}$. 
\item {\em Step 2}: if $Z_N^{(L)}\geq C_L$ but $Z_N^{(L-1)}<C_{L-1}$, reject $H_{(L)}$ and
accept $H_{(1)},\ldots,H_{(L-1)}$ 
\[\vdots\]
\item {\em Step l}: if $Z_N^{(L)}\geq C_L,\ldots, 
Z_N^{(L-l+2)}\geq C_{(L-l+2)}$, but $Z_N^{(L-l+1)}< C_{(L-l+1)}$ reject
$H_{(L)},\ldots,H_{(L-l+2)}$ and accept
$H_{(1)},\ldots,H_{(L-l+1)}$. 
\[\vdots\]
\item {\em Step L+1}: if $Z_N^{(L)}\geq C_L,\ldots, Z_N^{(1)}\geq C_{1}$,
reject  $H_1,\ldots,H_L.$ 
\end{itemize}

\begin{remark}
 For each of these tests $\hat{\boldsymbol{T}}_{0}(f) =\boldsymbol{{\hat S}}(f)$ and
$\hat{\bfT}_{1}^{-1}(f)$ has only a single zero constraint so that finding it does not require the iterative scheme in \cite{WermuthScheidt77}. Consequently, the test statistics may be assembled very easily and efficiently.
\end{remark}

\subsection{Critical Levels}\label{subsec:critlev}
The choice of the critical values $C_1,\ldots, C_L$ is related to the idea of the family-wise error rate (FWER). If $Y$ is the number of true null hypotheses that are falsely rejected, then the FWER is defined as
$P(Y \geq 1),$
i.e., the probability that at least one true null hypothesis will be falsely rejected.
It is desired that ${\rm FWER} \leq \alpha$ for all possible constellations of true and false hypotheses, the so-called strong error control \cite[(9.3)]{LehmannRomano05}.
This can be achieved using the (conservative) Holm approach \cite[p.~363]{LehmannRomano05}: at each level the critical value can be evaluated using
$
C_i(\alpha)= F^{-1}\left(1-\frac{\alpha}{i}\right),
$
where $F(\cdot)$ denotes the common distribution function of the test statistic under the null hypothesis, which from (\ref{eq:sn}) is in fact $\Phi(\cdot),$ the standard Gaussian distribution function, in our case.
So we choose our critical values according to the easily computed formula
\begin{equation}\label{eq:alphaMHT}
C_i(\alpha)= \Phi^{-1}\left(1-\frac{\alpha}{i}\right).
\end{equation}

\subsection{Worked Examples}
Using the same $\mbox{VAR}_{5}(1)$ observations  as in  Section~\ref{sec.workex},   we list our $L=10$ hypotheses:

$$\begin{array}{ll}
H_{1}: & \mbox{Edge doesn't exist between} \hspace{2mm} (1,2) \\
H_{2}: & \mbox{Edge doesn't exist between} \hspace{2mm} (1,3) \\
\vdots & \vdots\\
H_{10}: & \mbox{Edge doesn't exist between} \hspace{2mm} (4,5) \\
\end{array}$$
Ordering the test statistics and including the critical levels $C_{i}(0.05)$ of (\ref{eq:alphaMHT}) gives Table~\ref{tab:MHTEx}.
\begin{table}[t!]
\begin{center}
\begin{tabular}{|c|c|c|c|}
\hline 
$i$ & Missing Edge & $Z_{N}^{(i)}$ & $C_{i}(0.05)$\tabularnewline
\hline 
\hline 
10 & (4,5) & 73.17 &  2.58\\
9 & (1,5) & 67.92 & 2.54\\
8 & (1,2) & 53.71 & 2.50\\
7 & (1,4) & 22.25 & 2.45\\
6 & (2,4) & 18.16 & 2.39\\
5 & (1,3) & 12.72 & 2.33\\
4 & (3,5) & 5.86 & 2.24\\
3 & (2,5) & 1.89 & 2.13\\
2 & (2,3) & 0.54 & 1.96\\
1 & (3,4) & 0.21 & 1.64\\
\hline
\end{tabular}
\end{center}
\caption{Ordered  statistics $Z_N^{(i)}$ and critical levels $C_i(0.05)$ for MHT}
\label{tab:MHTEx}
\end{table}
We can see that $Z_{N}^{(10)} \geq C_{10}(0.05), \ldots, Z_{N}^{(4)} \geq C_{4}(0.05)$ and $Z_{N}^{(3)}<C_{3}(0.05)$, so we reject $H_{(10)} \dots H_{(4)}$ and accept $H_{(3)} \dots H_{(1)}$. 
Note that this means our estimated graphical model is the graph with edges $\{(2,5), (2,3), (3,4) \}$ missing, the true graph for the model. 

We also compared behaviours of Model~B of Section~\ref{subsec:Vars} using $x=0,$ with Model~C, the only parametric difference being that  $\bfSigma_{\bfsepsilon}\not=\bfI_5$ for Model~C. The former has
missing edges  $\{ (2,3), (2,5) \}$ the latter has only $(2,3)$ missing. Constructing a table like Table~\ref{tab:MHTEx} for each we find for Model~B that
edges (2,3) and (2,5) have associated statistics 1.49 and -0.31 and are classified as missing, all other hypotheses are rejected. For Model~C 
edge (2,3) has associated statistics 1.07  and is classified as missing, all other hypotheses are rejected. So again the true graphs were found.

\section{Efficiency Contrast}\label{sec:efficiency}
\begin{prop}
The number of test statistics calculated in the Matsuda algorithm is $O(p^4)$ and in the MHT is $O(p^2).$
\end{prop}
{\it Proof:\/}
For Matsuda's algorithm, assuming the final output is the true graphical model with $k$ missing edges, 
\begin{eqnarray} \label{eqn.totalStats}
\frac{p(p-1)}{2} &+& \left[ \frac{p(p-1)}{2} -1 \right ] + \dots +  \left[ \frac{p(p-1)}{2} - k \right ]\nonumber\\
&=& (k+1)\frac{p(p-1)}{2} - \frac{k(k-1)}{2}
\end{eqnarray}
test statistics are calculated, where $k \in \{0, \ldots, p(p-1)/2 \}$.
Setting the ratio of non-edges to total possible edges  to $a$, we can write $k= a\frac{p(p-1)}{2}$ for $0 \leq a \leq 1$. Then substituting into (\ref{eqn.totalStats}), the total number of test statistics needing to be calculated, $n$ say, satisfies
$$n = p^4 \left[\frac{a}{4} - \frac{a^2}{8}\right] + o(p^4),$$
where $o(p^4)$ denotes terms of smaller order than $p^4.$
For sparsity take $1/2<a<1,$  then asymptotically in $p,$ 
$$\frac{3p^4}{32}  <n<\frac{p^4}{8},$$
i.e., $O(p^4).$
For the MHT, regardless of the number of missing edges in the model, we always calculate $n=p(p-1)/2$ statistics, so asymptotically, $n \approx p^2 /2,$ i.e., $O(p^2)$. 
\hfill $\square$
\medskip

Clearly the sample size, $
N,$ and length of weight sequence, $2M+1,$ will affect the time it takes to calculate each test statistic. 
Also, if there is only one missing edge in our model, as is the case in the MHT, we do not need to iterate in order to find the matrix satsfying the constraints in (\ref{eqn.spectralestimate}). If there is more than one missing edge, as in all steps of the Matsuda algorithm excluding the first, iteration is required as set out in \cite{WermuthScheidt77}.
As the number of iterations must increase as more edges are removed from the model for a good estimate, we will denote this number at each stage as $l_k$. ($l_k= 1$ in the MHT as we only have to iterate once). It can be shown by considering the steps in the construction process that computation time for each statistic is $\sim 2NM + Np^2l_k$. 

Combining this with the number of test statistics needed to be calculated above, Matsuda's algorithm has a time $T_1\sim 2NMp^4 + Np^6l_k$ and for the MHT,  
\begin{equation}\label{eq:MHTtimea}
T_2\sim 2NMp^2 + Np^4.
\end{equation}

So the calculation times $T$ for the tests would be expected to be 
\begin{equation}
T=
\begin{cases}
O(p^6) & \text{for Matsuda's algorithm};\\
O(p^4) & \text{for the MHT}.
\end{cases}
\end{equation}

\section{Practical Comparison For  Small Dimensions}\label{sec:computetime}

For small values of $p$ we  are able to make direct practical comparisons of the two algorithms as Matsuda's can still be calculated in a reasonable time period.
\subsection{Timings}\label{subsec:timingsone}
\begin{figure}[t!]
\begin{center}
\includegraphics[scale=1]{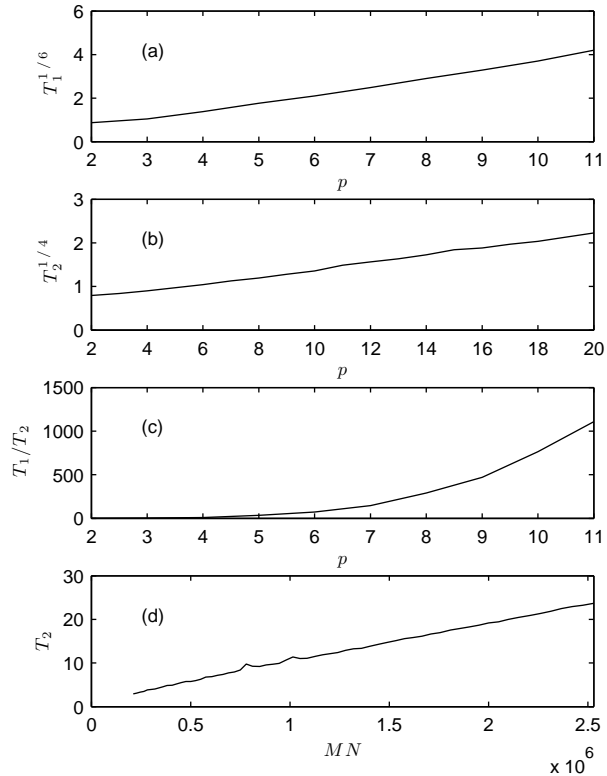}
\end{center}
\caption{Calculation timings in seconds:  (a) $T_1$ for Matsuda's algorithm, to the one-sixth power, versus $p$, (b) $T_2$ for the MHT, to the one-quarter power, versus $p$, 
(c) the ratio of computation times $T_1/T_2$ versus $p,$ and (d) $T_2$ for  the MHT versus $MN.$
Here $N=1024, M=32.$}
\label{fig:timings}
\end{figure}

Fig.~\ref{fig:timings} compares calculation times
$T$ in seconds, for the tests for $N=1024, M=32.$ 
Fig.~\ref{fig:timings}(a) plots $T_1^{1/6}$ versus $p$ for Matsuda's algorithm, while
Fig.~\ref{fig:timings}(b) plots $T_2^{1/4}$ versus $p$ for the MHT. In both plots these times increase linearly with $p$ as expected.
Fig.~\ref{fig:timings}(c) shows the ratio $T_1/T_2,$ illustrating the rapid increase in computation time for Matsuda's algorithm with $p,$ compared to the MHT approach. These results were derived by randomly generating a  $\varp(1)$ model matrix $\bfPhi_1$ (see the Appendix) for each $p$ value considered,  and then recording the completion time of each algorithm --- Matsuda's or MHT --- for that model. 

Fig.~\ref{fig:timings}(d) shows, for $p=5$ fixed and the MHT, a plot of $T_2$ versus $MN,$ where $M=N/32$ and $N$ increases from 200 to $9\,000.$  From
(\ref{eq:MHTtimea})
$$
T_2\sim 2NMp^2 + \frac{NMp^4}{M} \Rightarrow \frac{\dif T_2}{\dif (NM)}\sim 2p^2+ \frac{32 p^4}{N},
$$
which for large $N$ means that $T_2$ should have a constant gradient with $MN,$ as seen in Fig.~\ref{fig:timings}(d). These results were derived 
by randomly generating a single  ${\rm VAR}_5(1)$ model matrix $\bfPhi_1,$ (Appendix), and then recording the completion time for the MHT algorithm for that model
using the different $M, N$ combinations specified. 
%

\subsection{Power}\label{sec:power}
We will compare the results of the MHT approach against Matsuda's algorithm using two different models.
To do this we utilise the concepts of (i) FWER, defined in Section~\ref{subsec:critlev}, and (ii) effective power, the probability of rejecting all false hypotheses \cite{Shaffer95}.

The first model is the  VAR$_5(1)$ model of  (\ref{eq:MatPhi}) 
and we consider the cases $x=0$ (missing edges $\{ (2,3), (2,5) \}$), and $ 0.1,$ (single missing edge $\{(2,5) \}$), as used in \cite{Matsuda06}.

We considered combinations $(N, M)$ of $(512, 16),$ $(1024, 32),$ $(2048, 64)$. Results are based on 600 replications for each $(N, M)$ pair.

For $x=0$ to compare the algorithms, we only consider the edges $(2,3),(2,5),(3,4)$. This is due to the fact that these produce the three borderline statistics and while others may sometime fall outside the critical region --- i.e., we reject them as edges --- this is infrequent enough that simply for comparison purposes it is worth saving time by ignoring these. This approach is supported by the results in Table~\ref{tab:avsemodelone} which used the values $N = 2048$ and $M = 64$. (In the computations the test statistics for other edges were essentially taken to be infinity.)
\begin{table}[t!]
\begin{center}
\begin{tabular}{|c|c|c|}
\hline 
Edge & Average & Standard Error\tabularnewline
\hline 
\hline 
(1,2) & 26.93 & 4.57\tabularnewline
\hline 
(1,3) & 37.94 & 5.25 \tabularnewline
\hline 
(1,4) & 12.55 & 3.10\tabularnewline
\hline 
(1,5) & 41.39 & 5.63\tabularnewline
\hline 
{\bf (2,3)} & {\bf 0.25} & {\bf 1.08}\tabularnewline
\hline 
(2,4) & 33.21 & 5.03\tabularnewline
\hline 
{\bf (2,5)} & {\bf 0.34} & {\bf 1.05}\tabularnewline
\hline 
{\bf (3,4)} & {\bf 1.00} & {\bf 1.21}\tabularnewline
\hline 
(3,5) & 13.40  & 3.39\tabularnewline
\hline 
(4,5) & 15.39  & 3.68\tabularnewline
\hline 
\end{tabular}
\end{center}
\caption{Average and standard error of values of the Model~B ($x=0$) test statistic $Z_N^i$ for each edge test with $N=2048, M=64$.}
\label{tab:avsemodelone}
\end{table}

\begin{table}[t!]
\begin{center}
\begin{tabular}{|c|c|c|c|}
\hline 
\multicolumn{1}{|c|}{$x$} & \multicolumn{3}{|c|}{missing edge hypothesis}\tabularnewline
 & (2,3) & (2,5) & (3,4)\tabularnewline
\hline \hline 
0 & True & True & False \tabularnewline
\hline 
0.1 & False & True  & False \tabularnewline
\hline 
\end{tabular}
\end{center}
\caption{State of the missing edge hypotheses for Model~B when $x=0$ and $x=0.1.$ }
\label{tab:statehyp}
\end{table}

\begin{figure}[t!]
\begin{center}
\includegraphics[scale=1]{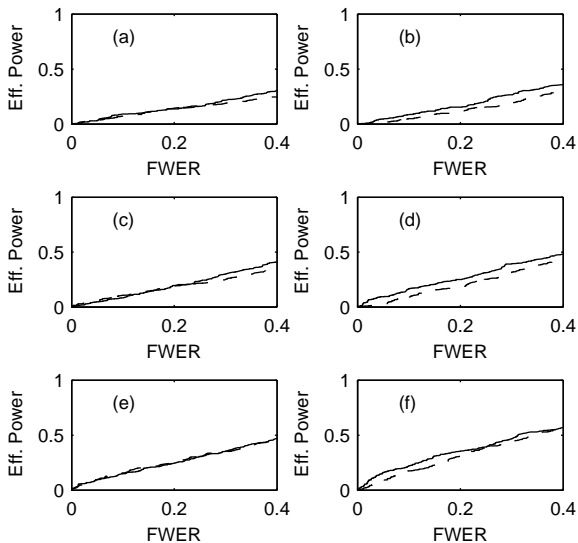}
\end{center}
\caption{FWER versus effective power for the MHT (solid lines) and Matsuda's algorithm (dashed line) for Model~B, (\ref{eq:MatPhi}), with (a) $N=512, M=16, x=0$ (b) $N=512, M=16, x=0.1,$ (c) $N=1024, M=32, x=0$ and (d) $N=1024, M=32, x=0.1,$ (e) $N=2048, M=64, x=0$ and (f) $N=2048, M=64, x=0.1.$}
\label{fig:FWER}
\end{figure}

The results displayed in Fig.~\ref{fig:FWER} were constructed as follows. 
For the multiple hypothesis test, 
$\alpha$ was varied between $0$ and $0.5$ in steps of $0.00125,$  and used as in (\ref{eq:alphaMHT}). The MHT was carried out for each of the 600 replications followed by the two steps:
\begin{enumerate}
\item
 the FWER was recorded as the proportion of the replications for which at least one true null hypothesis was falsely rejected; 
\item
the effective power of the test was recorded as 
the proportion of replications for which 
$(3,4)$ was not included as a missing edge. This is essentially the power of the sub-test on the hypotheses claiming edges $(2,3), (2,5), (3,4)$ to be missing, since of these the only  hypothesis that is false is the $(3,4)$ one; see Table~\ref{tab:statehyp}.
\end{enumerate}

For Matsuda's algorithm a parameter $\beta$ was created and varied between $0$ and $0.5$ in steps of $0.00125,$   and then $\alpha$ formed from $\alpha = \beta^{5};$  this $\alpha$ is the quantity used in (\ref{eq:alphaMat}).
This approach allowed us to concentrate more $\alpha$ values near zero, resulting in a more even grid for the resultant FWER. 
Matsuda's algorithm was carried out for each of the 600 replications and the FWER and effective power recorded.

\begin{figure}[t!]
\begin{center}
\includegraphics[scale=1]{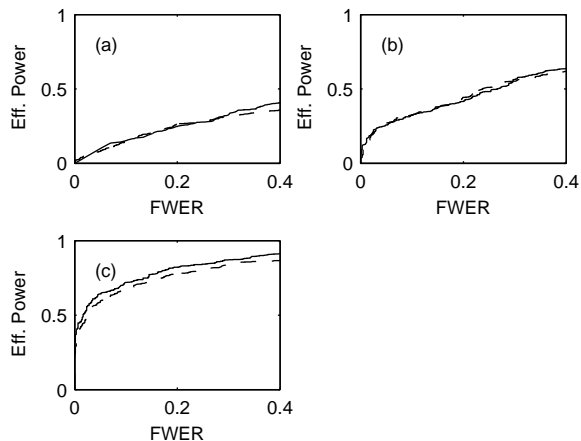}
\end{center}
\caption{FWER versus effective power for the MHT (solid lines) and Matsuda's algorithm (dashed line) for Model~A, (\ref{eq:modelone}),  and (a) $N=512, M=16$  (b) $N=1024, M=32$ and (c) $N=2048, M=64.$}
\label{fig:MM_FWER}
\end{figure}
Figs.~\ref{fig:FWER}(a), (c) and (e) show the relationship between the FWER and effective power for the MHT (solid line) and Matsuda's algorithm (dashed line). As can be seen, there is no significant difference in the power of the test for the two methods.

For the case $x=0.1$ we see from Table~\ref{tab:statehyp} that the hypotheses stating 
$(2,3)$ and $(3,4)$ to be missing edges are both false. So the same basic procedure is carried out as for $x=0$ but now the effective power is computed as the probability of rejecting both the hypotheses involving $(2,3)$ and $(3,4).$ The results are shown in Figs.~\ref{fig:FWER}(b), (d) and (f)
from which it is seen that again the MHT does at least as well as Matsuda's algorithm.

Turning to model~A
of Section~\ref{subsec:Vars}, with $\bfPhi_1$ given in (\ref{eq:modelone})
and missing edges  $\{(2,3),(2,5),(3,4)\},$ we can see in Table~\ref{tab:avsemodeltwo} that the only other `boundary edge' is $(3,5)$.
\begin{table}[t!]
\begin{center}
\begin{tabular}{|c|c|c|}
\hline 
Edge & Average & Standard Error\tabularnewline
\hline 
\hline 
(1,2) & 50.00 & 5.93\tabularnewline
\hline 
(1,3) & 15.74 & 3.52 \tabularnewline
\hline 
(1,4) & 22.02 & 4.34\tabularnewline
\hline 
(1,5) & 64.12 & 6.79\tabularnewline
\hline 
{\bf (2,3)} & {\bf 0.29} & {\bf 1.06}\tabularnewline
\hline 
(2,4) & 15.66 & 3.38\tabularnewline
\hline 
{\bf (2,5)} & {\bf 0.27} & {\bf 1.06}\tabularnewline
\hline 
{\bf (3,4)} & {\bf 0.32} & {\bf 1.05}\tabularnewline
\hline 
{\bf (3,5)} & {\bf 3.86}  & {\bf 1.95}\tabularnewline
\hline 
(4,5) & 66.09  & 6.61\tabularnewline
\hline 
\end{tabular}
\end{center}
\caption{Average and standard error of values of the model 2 test statistic $Z_N^i$ for each edge test with $N=2048$ and $M=64$.}
\label{tab:avsemodeltwo}
\end{table}

Again we considered combinations $(N, M)$ of $(512, 16),$ $(1024, 32),$ $(2048, 64)$ and used 600 replications for each $(N, M)$ pair. The results were calculated using the same method as above, the only difference being the effective power is now the power of the sub-test on hypotheses claiming the edges $(2,3),(2,5),(3,4),(3,5)$ to be missing. Of these, the false hypothesis is that stating $(3,5)$ to be a missing edge.

Fig.~\ref{fig:MM_FWER} compares the FWER and effective power for the MHT  and Matsuda's algorithm. Again, there is no significant difference in the power of the test for the two methods.

\section{MHT Algorithm For Higher Dimensions}\label{sec:highdims}
We have shown that the MHT approach performs well for a relatively small number of dimensions $p.$ We now look at higher dimensions.
\subsection{Timings}
\begin{figure}[t!]
\begin{center}
\includegraphics[scale=1]{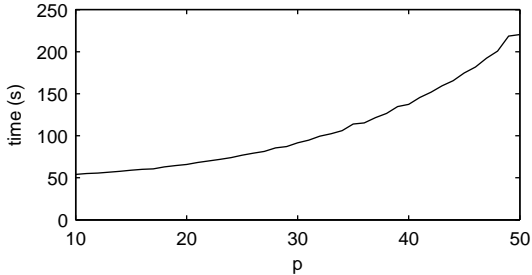}
\end{center}
\caption{Calculation timings in seconds for the MHT algorithm as $p$ varies from 10 to 50. Here  $N=2048$ 
and $M=128.$}
\label{fig:timebigp}
\end{figure}

It might be thought that the inefficiency of Matsuda's algorithm  is not of concern for such moderately large $p,$  given modern computing power. 
However Fig.~\ref{fig:timebigp} gives timings (see Section~\ref{subsec:timingsone}) for the MHT algorithm  in seconds for $p$ from 10 to 50 (using a  3GHz processor). Here $N=2048$ and $M=128.$  For $p=50$ the time taken was about 220s; if this is scaled up (crudely) for Matsuda's algorithm by $p^2=2500$ we arrive at a time of over 6 days.

\begin{table}[t!]
\begin{center}
\begin{tabular}{|c|c|c|c|}
\hline 
&$p=10:29$ & $p=30$ & $p=30:50$
\tabularnewline
\hline \hline 
Type I & 2.2 & 3.0 & 4.1 \tabularnewline
\hline 
Type II & 1.3 & 2.4  & 2.9 \tabularnewline
\hline 
\end{tabular}
\end{center}
\caption{Average  type I and II percentage errors }
\label{tab:percent}
\end{table}

\subsection{Accuracy}
Table~\ref{tab:percent} reports the average type I and type II percentage errors encountered in the model estimation when $\alpha=0.05.$
Here averaging is (i) over the 20 estimated models for $p=10:29$ (first column), (ii) over 100 repeat simulations for the single case $p=30$ (second column), and (iii) over the 21 estimated models for $p=30:50$ (third column). The type I percentage error is here the ratio 100(number of edges accepted when missing)/(number missing) and the type II percentage error is the ratio 100(number of edges deleted when present in the true graph)/(number present).

\begin{figure}[t!]
\begin{center}
\includegraphics[scale=1]{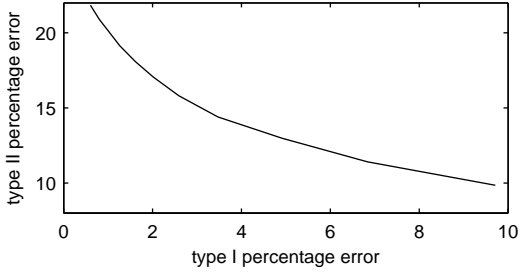}
\end{center}
\caption{Type I and II percentage errors 
for $p=150$ as $\alpha$ is varied. Here $N=2048, M=512$.
}
\label{fig:type1and2errorsp150}
\end{figure}

Fig.~\ref{fig:type1and2errorsp150} gives the type I and II percentage errors 
when $p$ is fixed at the large value $p=150$ and $\alpha$ is varied. 
These results were derived using a  $\varp(1)$ model matrix $\bfPhi_1$ (see the Appendix) giving rise to a true graphical model with 36\% of connections present. The results seem quite satisfactory and behave in the reciprocal way expected. 

\subsection{Parallelizability}
\begin{figure}[h]
\begin{center}
\includegraphics[scale=1]{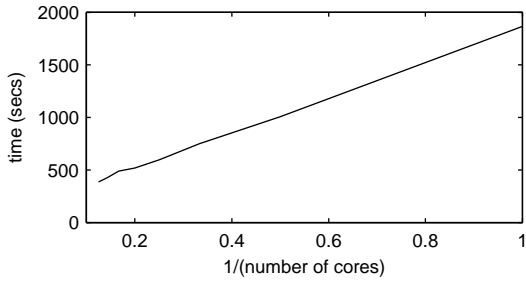}
\end{center}
\caption{Calculation timings in seconds for the MHT algorithm for $p=60, N=2048, M=512$ against the reciprocal number of cores, as the number of cores varies from 1 (right of plot) to 8 (left).}
\label{fig:parallel}
\end{figure}

In contrast to Matsuda's implementation there is no dependency between the calculation of each of the test statistics. On a multicore CPU a test statistic can be assigned to each core, and upon completion  the next statistic needing calculation is assigned. Due to the small overheads this introduces, compute-time minus a (near) constant factor for calculating the frequency-averaged periodogram is simply  inversely proportional to the number of cores used. 
The fact that for large $p$ most time is spent in calculating the test statistics means that our algorithm can be effectively scaled for higher dimensionality just by using more processor cores.
The example of Fig.~\ref{fig:parallel} illustrates the inverse proportionality, using an 8 core  processor.


\section{Application to EEG Data}\label{sec:EEG}
We now apply the MHT method
to electroencephalogram (EEG) data, (resting conditions with eyes closed), for 33 males, 19 diagnosed with  negative-syndrome schizophrenia, and 24 controls. 
This rare heritage clinical dataset from unmedicated patients  was discussed in detail in \cite{Medkouretal10}. Interest is in detecting any differences in patterns of brain connectivity between the groups.

\begin{figure}[t]
\begin{center}
\includegraphics[scale=1]{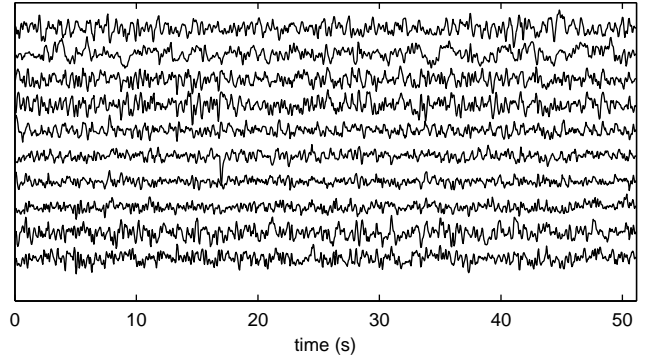}
\end{center}
\caption{Ten channel EEG time series for one of the negative-syndrome patients.}
\label{fig:plot_ten_neg_indv3}
\end{figure}
\begin{figure}[t]
\begin{center}
\includegraphics[scale=1]{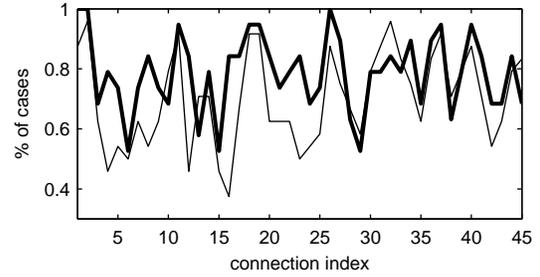}
\end{center}
\caption{Percentage of negative-syndrome patients (heavy line) and controls (thin line) exhibiting a specified connection. ($N=1024, M=20, \alpha=0.01).$}
\label{fig:controls_and_negative}
\end{figure}
For each individual
EEG was recorded on the scalp  at $10$  sites so that $\{\bfX_t\}$ is a $p=10$ vector-valued process.
 There are
$2^{p(p-1)/2}=2^{45}$ possible graph structures, 
and $p(p-1)/2=45$ possible connections between the series (edges to the graph). Each possible connection was assigned a connection index from 1 to 45 as given in \cite{Medkouretal10}.

For illustration purposes, the ten channel time series for one of the 
negative-syndrome patients is shown in 
Fig.~\ref{fig:plot_ten_neg_indv3}.
For each of the negative-syndrome patients
the MHT algorithm was used to determine whether an index-$i$
connection existed, and the percentage of the group of patients exhibiting this connection was recorded. The same was done for the control group.  Fig.~\ref{fig:controls_and_negative}
gives the resulting percentages for each connection and both groups. For 3/4 of the connections the percentage is lower for the controls, suggesting patients exhibit a tendency towards higher 
connectivity, a result consistent with  \cite{Medkouretal10} where completely different methodology was used.

\section{Concluding Discussion}\label{sec:summary}

Matsuda's approach to identification of a graphical model involves an appealing Kullback-Leibler statistic but,
while improving on exhaustive search approaches, 
his implementation using a backward stepwise selection is extremely heavy computationally.
This paper introduced a multiple hypothesis test implementation using Matsuda's statistic. The number of statistics needing to be calculated is reduced by $O(p^2)$ and the computational burden for evaluating the test
statistics themselves is notably reduced as iterative fitting algorithms are no longer required. 

The MHT approach allows us to derive a more relevant control on the error rate in contrast to the stepwise procedure where the error rate used in each test step doesn't have a clear link to the total error of the procedure.
The type I error rate we are controlling is the probability of failing to delete an edge when it is missing in the true graphical model. It may be more intuitive to define the error as  deleting an edge that is contained in the true graph. In order to do this we would have to accurately know the distribution of the test statistic under this alternative, which unfortunately we don't know this. 

The conservative nature of the Holm approach can in theory be somewhat offset by using an adaptive approach, (explained in detail by Guo \cite{Guo09}), particularly for large $p.$ The result is a more powerful test than the standard Holm procedure and although the FWER will be higher, Guo showed it still controls the FWER asymptotically. We implemented this methodology but for our examples and the values of $p$ utilised, differences were very small; however, this approach is undoubtedly worthy of further investigation.


It is possible using our method to conduct an efficient stepwise approach by running the MHT and keeping all edges that clearly exist (i.e., have a very large test statistic), thus defining a new $\bfT_0$ to that used previously. Much of the work is thus completed. Then the MHT can  be re-run to test models differing from $\bfT_0$ by one edge, but such additional steps  require the iterative scheme  \cite{WermuthScheidt77}. 

Finally, we have shown that the algorithm scales very well --- is highly parallelizable --- with appropriate computing resources. Future work would involve rendering the algorithm for efficient calculation on high performance computing hardware such as GPUs.

\subsection*{Appendix: Random Model Construction}
\label{app:randmod}
For our simulations random $\varp(1)$ models were constructed by randomly formulating  $p\times p$ matrices $\bfPhi_1$ with the number of zero entries specified as follows.

For a given $p$ value a $p \times p$ matrix $\bfPhi_1$ was constructed with null entries. All diagonal elements and non-diagonal elements in position $(i,j)$ for which $(i+j)_{\!\!\!\mod k}=1$
were populated by random values sampled from the ${\cal N}(0,1)$ distribution. The matrix was then subject to spectral decomposition and any eigenvalues with modulus greater than unity were replaced by their reciprocals and $\bfPhi_1$  reconstructed using the modified eigenvalues. 
For such a $\bfPhi_1$ we know 
$
\det\{ {\bfI}_p-\bfPhi_1 z \}\not=0\,\,\mbox{for all}\,\, |z|\leq 1,
$
\cite[pp.~15 \& 653]{Lutkepohl06} and so a stationary process results. The choice of $k$ controls the sparsity; our default choice  $k=5$ makes approximately $64$\% of the
$\bfPhi_1$ matrix entries zero for $p=10:50.$

\section*{Acknowledgement}
The work of Rob Wolstenholme was supported by EPSRC (UK). 
.

\nocite{*}
\bibliographystyle{IEEE}

\end{document}